# 250 Magnetic Tunnel Junctions-Based Probabilistic Ising Machine

Shuhan Yang[1,5], Andrea Grimaldi[2,5], Youwei Bao[1], Eleonora Raimondo[3,2], Jia Si[4], Giovanni Finocchio[2*] and Hyunsoo Yang[1*]

[1]*Department of Electrical and Computer Engineering, National University of Singapore, Singapore.*

[2]*Department of Mathematical and Computer Sciences, Physical Sciences and Earth Sciences, University of Messina, Messina, Italy.*

[3]*Istituto Nazionale di Geofisica e Vulcanologia, Rome, Italy.*

[4]*Key Laboratory for the Physics and Chemistry of Nanodevices and Center for Carbon-based Electronics, School of Electronics, Peking University, Beijing 100871, China.*

[5]*These authors contributed equally*

*e-mail: giovanni.finocchio@unime.it, eleyang@nus.edu.sg

**In combinatorial optimization, probabilistic Ising machines (PIMs) have gained significant attention for their acceleration of Monte Carlo sampling with the potential to reduce time-to-solution in finding approximate ground states. However, to be viable in real applications, further improvements in scalability and energy efficiency are necessary. One of the promising paths toward achieving this objective is the development of a co-design approach combining different technology layers including device, circuits and algorithms. Here, we experimentally demonstrate a fully connected PIM architecture based on 250 spin-transfer torque magnetic tunnel junctions (STT-MTJs), interfaced with an FPGA. Our computing approach integrates STT-MTJ-based tunable true random number generators with advanced annealing techniques, enabling the solution of problems with any topology and size. For sparsely connected graphs, the massive parallel architecture of our PIM enables a cluster parallel update method that overcomes the serial limitations of Gibbs sampling, leading to a 10 times acceleration without hardware changes. Furthermore, we prove experimentally that the simulated quantum annealing boosts solution quality 20 times over conventional simulated annealing while also increasing robustness to MTJ variability. Short pulse switching measurements indicate that**

**STT-MTJ-based PIMs can potentially be 10 times faster and 10 times more energy-efficient than graphic processing units, which paves the way for future large-scale, high-performance, and energy-efficient unconventional computing hardware implementations.**

The rapid advancement of artificial intelligence and the Internet of Things intensifies the demand for solving problems with high computational complexity, which are critical for applications such as data routing, topology optimization, etc. Non-deterministic polynomial time hard (NP-hard) problems are the most computationally challenging ones, requiring exponentially increasing resources with the growth of the problem size[1]. Practical applications of NP-hard problems span various fields in the form of combinatorial optimization problems (COPs)[2] such as scheduling[3], resource allocation[4], logistics[5], and financial management[6,7].

As Moore's Law approaches its limits[8], interest is growing in unconventional hardware solution to tackle complex COPs with larger energy efficiency and computational speed[9-17]. Probabilistic Ising Machines (PIMs) are emerging as promising candidates for improving both factors[18-22]. Several PIM implementations leveraging a variety of technology platforms have been proposed, including memristors[23,24], optics[25], ferroelectric transistors[26,27], fully digital field programmable gate arrays (FPGA)[19], and others[28,29]. Considering that a main ingredient of the PIMs is the massive generation of random numbers, hybrid solutions have emerged that integrate technologies highly effective for random number generation with semiconductor technology. In particular, magnetic tunnel junctions (MTJs) stand out as promising spintronic candidates due to their energy and area efficiency, driven by their inherent stochasticity in the switching process and tunable switching probabilities[30]. Specifically, superparamagnetic tunnel junction (SMTJ)-based PIMs have been widely investigated in both theoretical studies[31-33] and hardware prototypes, where SMTJs have been used as probabilistic bits[34-36] and as asynchronous clocks[37]. While SMTJs offer energy efficiency in random number generation due to the low energy barrier, the thermal fluctuation-driven switching makes the system susceptible to environmental changes. Because of the lack of native data retention, it places strict requirements on peripheral reading circuits, which must capture the states of the fast-switching nanomagnets quickly enough before they change states. Furthermore, the small volume of the free layer makes controlling device-to-device fluctuations in large-scale PIMs difficult, which poses a major challenge in scaling. STT-MTJs also exhibit stochastic switching behaviour with appropriate input pulses, making them suitable for random number generation[38]. However, unlike SMTJs, STT-MTJs possess a larger energy barrier between two stable states, providing greater immunity to environmental fluctuations. In addition, their non-volatile nature reduces the burden on peripheral circuits. As key elements of magnetic random-access memory (MRAM) technology, STT-MTJs provide superior uniformity compared to SMTJs. Previous studies have demonstrated probabilistic computing using a single STT-MTJ device[39] and a three-device STT-MTJs system[40]. However, to make STT-MTJ-based PIM viable for solving real-world optimization problems, further investigation into system-level scaling and algorithm implementation is essential.

This article presents a proof-of-concept PIM that integrates 250 STT-MTJs with an FPGA. The key contributions are: 1) demonstration of a thermally stable, non-volatile STT-MTJ PIM whose superior uniformity over SMTJs enables scalable implementation; 2) proof-of-concept operation at the 250-device level with evaluation of algorithmic performance under real device variations; and 3) a massively parallel, reconfigurable architecture that can simultaneously drive all or subsets of STT-MTJs, supporting multi-replica Gibbs sampling in parallel for fully connected problems, parallel cluster updates for sparse problems, and advanced energy-minimization techniques such as parallel tempering and simulated quantum annealing, which facilitates information exchange among different replicas of the problem. The architecture allows users to select algorithms according to the required trade-off between speed and solution quality. Leveraging intrinsic STT-MTJ randomness and these algorithms, the system solves computationally challenging tasks, including exact problems like 24-bit integer factorization and NP-hard problems like MaxCut.

**Operation principle of PIM**

The solving strategy of Ising machines consists of mapping a COP in an Ising Hamiltonian and then its solution is connected to the ground state of this Hamiltonian. In a lattice with $N$ spins, where each spin is in states $s = \{s_1,...,s_N\}$, the Ising Hamiltonian ($\mathcal{H}$) is defined as

$$\mathcal{H}(s) = -\sum_{i<j}^{N} J_{ij} s_i s_j - \sum_{i}^{N} h_i s_i \tag{1}$$

where $s_i$ and $s_j$ represent the states of the $i^{th}$ and $j^{th}$ spins, which can be either +1 or −1. Each $i^{th}$ spin is biased by a local field $h_i$ and interacts with the $j^{th}$ spin through a coupling $J_{ij}$.

The fundamental unit of Ising machine, the spin, corresponds to the probabilistic bit (p-bit) in PIMs.[31] Its state can be updated using

$$s_i = sgn(rand(-1,+1) + tanh(I_i(s))) \tag{2}$$

where *rand* samples uniform random numbers in the interval [−1,+1], *sgn* is the sign function, and *tanh* is the hyperbolic tangent function. The input signal $I_i(s)$ received by the $i^{th}$ p-bit, which carries spin information from all other p-bits in the system, is updated by[31]

$$I_i(s) = \beta\left(h_i + \sum_{j}^{N} J_{ij} s_j\right) \tag{3}$$

where $\beta$ serves as an inverse temperature hyperparameter, regulating the annealing process. By iteratively updating Eq. (2) and (3), the system dynamically evolves toward lower energy configuration eventually reaching the optimal solution.

**STT-MTJ based PIM**

An MTJ device consists of two ferromagnetic layers: one acts as the reference layer and the other as the free layer. These two layers are separated by an MgO insulating layer that enables quantum tunnelling. Fig. 1a presents the tunnelling magnetoresistance (TMR) as a function of the voltage across the MTJ device, with a TMR ratio of approximately 180%. As depicted in the insets of Fig. 1a, between the stable antiparallel (AP, marked with A) and parallel (P, marked with C) states, there are unstable states (marked with B), which play a crucial role in defining the switching probability for a given duration and amplitude of current pulses. The basic structure of p-bit cell consists of 1 transistor and 1 MTJ as shown in the left inset of Fig. 1b. The blue symbols in Fig. 1b depict the sigmoid-like switching probability response to input pulses with an amplitude of $V_{in}$, and pulse width of 10 μs (see details in Methods). The switching probability is evaluated averaging over 10,000 input pulses. The right insets of Fig. 1b show the output voltage of the MTJ unit cell ($V_{out}$) as a function of measurement sequence for the input pulses at point A, B, and C, respectively, mirroring the letters in Fig. 1a. As $V_{in}$ increases, the MTJ transits from favouring the AP state, to having AP and P equally favoured, to finally having the P state favoured.

By integrating 250 STT-MTJs with a peripheral circuit and an FPGA, we develop a 250-node PIM (see also Supplementary Note 1). The p-bit behavior of Eq. (2) is achieved using the intrinsic randomness of STT-MTJ switching as shown in Fig. 1b. Fig. 1c illustrates the diagram of the system and Fig. 1d shows a photograph of the printed circuit board (PCB) and the FPGA. The system comprises 16 processing elements (PEs), each containing 16 MTJ computing units. Each MTJ computing unit includes an N-type metal–oxide–semiconductor (NMOS) transistor to regulate the current flowing through the MTJ device. 16 digital-to-analog converters (DACs) offer 256 analog input channels to the MTJ cells to control the switching probability, and 16 analog-to-digital converters (ADCs) provide 256 channels to sample the state of each MTJ cell after the perturb operation, thus allowing to individually access each MTJ cell. The use of ADCs in this work could be replaced with read sense amplifiers[36], that have been extensively developed for their use in MRAM technology, offering much smaller footprint and lower energy consumption. Fig. 1e shows the original stochastic switching responses of 250 MTJs, where noticeable deviations among devices exist. Through a linear transformation of the input signal (see also supplementary Note 2 and 3), all curves fall into a standard sigmoid shape in Fig. 1f. The value of the input $I$ is determined by Eq. (3) in our case.

An Ising solver can easily be trapped in local minimum states during the evolution process. Algorithms such as simulated annealing (SA)[41] and parallel tempering (PT)[42] are commonly used to reduce the probability of such a trapping. SA, for example, starts the system evolution at a high pseudo-temperature (low $\beta$) to overcome local energy minima and gradually cools down to a low temperature (large $\beta$) to reach the global minimum. The convergence and relaxation properties of SA have been rigorously proven mathematically[43]. In our PIM, SA requires random numbers in each iteration to generate new p-bit states, with true randomness being more effective in preventing local minima trapping. Low-quality random number generators (RNGs) can introduce systematic errors, deviating the system from the target Boltzmann distribution[22,37]. Our STT-MTJ-based PIM leverages the intrinsic stochastic switching of MTJ devices to achieve true randomness, minimizing hardware requirements, energy consumption, and time delays[44,45].

**Sequential, cluster parallel updating schemes and simulated quantum annealing.**

To ensure the proper evolution of the Markov chain, conventional Gibbs sampling requires every p-bit in a system to be updated sequentially so that one p-bit can capture the most recent state information from all the other p-bits. Therefore, having *n* p-bits generating one state each or one p-bit generating *n* states is mathematically equivalent, provided the input signal is properly calculated depending on which spin is currently being updated. Thus, in a sequential update scheme, a single MTJ cell (single p-bit) can update all spins without additional speed or energy loss. With 250 MTJs available, multiple independent solving attempts can be run in parallel. Each replica evolves independently according to its own sampling dynamics, and the best solution is selected based on the final energy evaluation. However, this serial update process significantly slows down the evolution of the system, especially as the number of spins increases. If the update of the p-bits is marked by a clock signal, each clock cycle would only comprise of one update. This is schematically shown in Fig. 2a. Here, the graph for a 10-bit integer factorization problem is illustrated in a circular layout. With the sequential update scheme, the 80 p-bit system would require 80 clock cycles to be updated once. As a single MTJ can be used to perform RNG for all the spins of the graph, 250 replicas of the same problem, each managed by an MTJ, can be run in parallel in our system.

For graphs that are not fully connected, which is common for many real-life problems[46]**Error! Reference source not found.**, p-bits that are not directly interacting can be updated simultaneously without disrupting the Markov chain, allowing for potential speedup in system. In the cluster parallel update scheme, we employ a greedy graph coloring algorithm to partition the p-bits into clusters, where each cluster is an independent set. This enables the simultaneous update of all p-bits within the same color. Although coloring is an NP-hard problem and finding the optimal solution can be a challenging task, our goal here is simply to divide the graph into independent sets. Therefore, a sub-optimal coloring strategy is suffice. Fig. 2b presents the coloring strategy for the 10-bit integer factorization problem, which is encoded with 80 p-bits and divided into 5 colors. The 250 MTJs available are divided into 15 replicas of 16 MTJs each. Each replica processes the colors in sequence, updating up to 16 p-bits at the same time. In an ideal case, this method offers substantial speed increase over the sequential update schemes of up to a factor of $N/G$, where $N$ is the total number of spins, and $G$ is the number of colors. We implement both sequential and parallel update methods on the Max-Cut problem. In terms of solution quality, both methods perform equally well (see details in Supplementary Note 4).

The other fundamental component of the algorithm is the energy minimization schedule used. Each schedule can be used with both sequential and cluster parallel update schemes. While SA is a robust energy minimization schedule, we also implement simulated quantum annealing (SQA) to evaluate and compare its performance with SA (see Methods for details). SQA is particularly well suited to architectures in which several replicas can be operated in parallel using building blocks that present device-to-device variations. SQA works by employing multiple replicas at the same temperature interacting through a transverse field that drives the superposition of all possible states into a final collapse to a ground state and becomes increasingly strong as the schedule progresses. As illustrated in Fig. 2c, in our PIM implementation, we use 15 sets of 16 SQA replicas and assign one MTJ to one SQA replica. Each replica follows the sequential updating scheme.

**Exact factorization of a 24-bit integer**

We benchmark our PIM on integer factorization problems, which are widely used in cryptographic applications, as the security of many cryptographic protocols relies on the difficulty of determining the prime factors of a semiprime. Unlike optimization problems, integer factorization is an exact problem where only factors that yield the correct product are considered valid solutions; near optimal or approximate solutions are not acceptable. Moreover, if the number to factor is a semiprime, then there is only a single non-trivial solution. While factoring a semiprime into its prime components is challenging, verifying the result by multiplying the factors is straightforward.

To implement integer factorization in our PIM (details on the encoding can be found in Supplementary Note 5), we use an SA process that begins with an inverse temperature $\beta = 0$, representing infinite temperature, allowing the system to explore a wide range of states. $\beta$ is then increased linearly until the temperature is low enough to freeze the spins into an energy minimum. Fig. 3a presents the annealing process of a 24-bit integer factorization (11,970,307) showing the normalized energy as a function of the inverse temperature $\beta$. The normalized energy is defined as $(E - E_{gs})/|E_{gs}|$, where $E$ represents the energy of a state, and $|E_{gs}|$ represents the absolute value of the ground state energy. When this value reaches 0, the system attains its ground state, indicating a successful factorization. It can be observed that the ground state is reached at high $\beta$ value (low temperature). The inset of Fig. 3a presents the solution cost versus the inverse temperature $\beta$, where solution cost is defined as $|F-AB|$, with $A$ and $B$ being the factors currently visited by the PIM and $F$ the number to factor. The solution cost reaches 0 when the integer is successfully factored. The histograms of factors $A$ and $B$, and of the product $F$ over the entire annealing process are shown in Fig. 3b, showing that the correct factors 3673 and 3259 are reliably visited (Supplementary Note 6), as well as the correct product of the two factors.

We implement both sequential and cluster parallel update schemes for integer factorization problems up to 24-bit (444 p-bits). For each instance, we conduct 1000 trials using the same annealing schedule to evaluate the time-to-solution (TTS) and iterations-to-solution. The TTS of a specific annealing schedule $\alpha$ is $TTS(\alpha) = T_\alpha \frac{\log(0.01)}{\log(1-P_\alpha)}$ where $T_\alpha$ is the time required to run the annealing schedule, and $P_\alpha$ is the probability of success evaluated among all the runs of the experiment. Similarly, the iterations-to-solution metric measures the same quantity as the TTS but scaled by number of iterations ($Z_\alpha$) of the annealing schedule. As shown in Fig. 3c, the iterations-to-solution are similar for both sequential and cluster parallel updating schemes, indicating that the two algorithms, as expected, perform equally well in solving integer factorization tasks. However, Fig. 3d demonstrates that the cluster parallel updating method achieves one order of magnitude reduction in TTS due to its parallel updating scheme. Despite the cluster parallel method requiring 16 times more hardware effort, the energy to solution does not increase because of the fewer updates needed for each MTJ, as illustrated in Fig. 3e. Therefore, the cluster parallel updating scheme is an effective way to accelerate the evolution process for not fully connected graphs. The more hardware redundancy available, the greater the acceleration achieved, although this advantage can only be leveraged to its full potential if the size of the instance increases accordingly.

**Comparison of replicated SA, parallel tempering and SQA on Max-Cut**

Compared to SA, which works well even with a single replica, parallel tempering (PT) and SQA requires several interacting instances running in parallel, causing it to be naturally more hardware intensive. With the same hardware effort, we can perform the number of replicas ($R$) independent SA solving attempts, so it is necessary to determine the SQA performance given the same allocation of resources. To compare the performance of replicated simulated annealing (replicated SA) with PT and SQA on our PIM, we benchmark them on Max-Cut problems using the Biq Mac dataset[47]. For replicated SA, we implement 16 replicas running in parallel and select the best solution at the end of the annealing process, ensuring equal hardware effort for replicated SA, PT and SQA (16 MTJ cells). We conduct 50 experiments for each algorithm, each experiment consisting of 10,000 iterations with optimized parameters, to compare the obtained maximum cut values.

Fig. 4 presents the comparison results of replicated SA, PT and SQA on five problems with different topologies, densities and solving difficulty (see details in Supplementary Note 7). More details on PT are outlined in the Methods section. Due to the large differences in optimal cut value of these instances, we define the approximation accuracy as the achieved cut value divided by the optimal cut value. 100% approximation accuracy represents reaching the optimal cut value. With the same number of resources allocated, SQA consistently outperforms both PT and replicated SA, achieving the highest median approximation accuracies and the smallest run-to-run variability across all five instances. PT trails closely behind SQA on the easier problems but suffers a larger drop on the hardest case (t2g20_5555). Replicated SA shows the greatest sensitivity to problem difficulty. The performance gap is particularly evident in large instances with large interaction strength and high dimensions. The 2 dimensional 200-node problem t2g20_5555, where the best trial using replicated SA performs worse than the worst trial using SQA. To further assess the superiority of SQA over conventional replicated SA, we evaluate the "superiority factor" (Supplementary Note 7), which establishes how many replicated SA runs are needed to match SQA. For 100-node problems, we observe the superiority factor of up to 20 times, which indicates that SQA is 20 times more efficient than replicated SA. The above result demonstrates that the interactions between replicas established by SQA (see Methods for details) can effectively increase the overall solution quality.

Another benefit of SQA is that it further alleviates the burden of calibrating large PIM systems. The interaction between replicas acts as an "error correction" technique that mitigates variations between devices. Moreover, the implementation of a complex algorithm like SQA demonstrates that our configurable PIM serves as an effective platform for implementing advanced annealing algorithms to improve solution quality.

**Benchmark with state-of-the-art PIMs**

We benchmark our STT-MTJ-based PIM with state-of-the-art probabilistic machine implementations. At device level, shown in Fig. 5, STT-MTJs outperform other technology platforms, including bistable resistor[48], memristor[23,22], FeFET[26], and SiOx nanorod[49]. The system level comparison (see also Supplementary Note 8) includes the NVIDIA V100 graphic processing unit (GPU)[50], FPGA[19], SMTJ[36], and SMTJ augmented FPGA[37] as shown in Fig. 5. Key metrics considered are the number of spin flips per second (FPS), and the energy consumed to generate one p-bit. A good probabilistic machine should exhibit high FPS and low energy consumption, positioning it in the lower right corner of Fig. 5. At the system level, our STT-MTJ-based PIM outperforms SMTJ and SMTJ-augmented FPGA implementations by achieving higher FPS with lower energy consumption.

From a speed perspective, the fastest experimentally demonstrated SMTJ has a retention time of 8 ns[51], which is slower than STT-MTJs, in which switching can be achieved with pulses as short as 1~2 ns[52,53]. In addition, SMTJs face challenges in yield control and device-to-device variation, which hinders large-scale implementations.

While a performance gap still exists between our system and mature full digital implementations such as FPGA and GPU, as shown in Fig. 5, our system generates true random numbers, whereas digital systems generate pseudo random numbers. The performance gap is primarily due to MTJ array size and circuit optimization, which can be bridged with MRAM-based application-specific integrated circuit (ASIC) design. Grounded by the 4 ns pulse switching data in a STT-MTJ (Supplementary Note 9), we project ~$10^{15}$ FPS and ~$10^{-13}$ J energy consumption per p-bit generation with a 1 Mbit array (see also Supplementary Note 8), which is ~10 times faster and 10 times more energy-efficient than GPU.

**Methods**

**Sample growth and device fabrication.** Thin film samples of substrate/[W (3)/Ru (10)]$_2$/W (3)/Pt (3)/Co (0.25)/Pt (0.2)/[Co (0.25)/Pt (0.5)]$_5$/Co (0.6)/Ru (0.85)/Co (0.6)/Pt (0.2)/Co (0.3)/Pt (0.2)/Co (0.5)/W (0.3)/CoFeB (0.9)/MgO (0.85)/CoFeB (1.2)/W (0.4)/CoFeB (0.8)/MgO (0.7)/Ta (3)/Ru (7)/Ta (5) were deposited via direct current (DC) (metallic layers) and radio frequency (RF) magnetron (MgO layer) sputtering on the Si substrates with thermal oxide of 300 nm with a base pressure of less than $2 \times 10^{-8}$ Torr at room temperature. The numbers in parentheses are thicknesses in nanometers. To fabricate the MTJs, bottom electrode structures with 10 μm width were firstly patterned via photolithography and Ar ion milling. MTJ pillars of a diameter of ~80 nm were patterned by using e-beam lithography. The encapsulation layer of Si$_3$N$_4$ was in-situ deposited after ion milling without breaking vacuum by using RF magnetron sputtering, and the top electrode structures with 10 μm width were patterned via photolithography, while the top electrode layers of Ta (5 nm)/Cu (40 nm) were deposited by using DC magnetron sputtering.

**PCB design.** Sixteen 16-channel DACs (AD5767) are used to generate analog input to each MTJ (through NMOS 2N7002), one 16-channel DAC (AD5767) generates bipolar $V_{dd}$ signals (16 MTJs share 1 $V_{dd}$), and sixteen 16-channel ADCs (MAX11131) are used to sample the output of each MTJ and feedback to FPGA (NI-SBRIO9651) for matrix multiplication. The amplitude of $V_{in}$ and $V_{dd}$ determine the current amplitude through the MTJ. The polarity of $V_{dd}$ dictates the current direction such that the reset and perturb processes can be realized. One 16-channel bipolar DAC is utilized to provide the $V_{dd}$ signal; thus all 16 MTJ cells of a processing element (PE) share one $V_{dd}$. Both DACs and ADCs interface with the FPGA through serial peripheral interface (SPI) protocol. The FPGA is programmed and controlled by LabVIEW. In our experiment, the main frequency of the PIM is set to 12.5 kHz. Considering the parallelism of 250 MTJs, the system generates 3.125 million spin flips per second.

**MTJ based p-bit characterization.** To generate one bit from the STT-MTJ-based p-bit, we first apply a reset pulse with negative $V_{dd}$ and zero $V_{in}$ to set the MTJ to the AP state. Following this, we zero-bias the MTJ cell and wait for the matrix multiplication to be conducted inside FPGA, which determines the value of $V_{in}$ for the perturb process. Finally, we apply a perturb pulse with positive $V_{dd}$ and positive $V_{in}$ to activate the stochastic behaviour of the MTJ (see Fig. 1b). The ADC then samples the output voltage $V_{out}$ and feeds this data back to the FPGA. Based on the pre-characterized threshold voltage ($V_{th}$) stored in the FPGA, if $V_{out} > V_{th}$, the current p-bit value is set to '+1', otherwise the value will be set to '−1'. Due to the hysteresis property of MTJ devices, a reset pulse is required before each random number generation. While in principle this transition could also be made stochastic by tuning the reset pulse amplitude (Supplementary Note 10), we only utilize the switching from the AP to P state to sample random numbers throughout this article. We use the deterministic switching from P to AP as the reset process and the stochastic switching from AP to P as the perturb process.

**Markov chain Monte Carlo (MCMC) and Gibbs sampling.** When considering a system with a large number of dimensions, finding the ground state analytically is often impossible. One way to investigate these systems is to employ a sampler that explores that state space of the problem by iteratively updating the values of its coordinates according to the energy functional. Differently from standard Monte Carlo, in which each sample is statistically independent from the others, MCMC samples are auto correlated. This correlation is fundamental to ensure the system samples from the chosen energy landscape. Among the many possible MCMC methods, Gibbs sampling is particularly appropriate when it is easy to calculate the conditional distribution of a variable, i.e., the distribution a specific variable has based on the state of all other variables. In the case of the Ising model, this conditional distribution is what we define as "input signal", Eq. (3).

**Parallel tempering.** Parallel tempering (PT) is an energy minimization algorithm that relies on parallel replicas of the system running in parallel at different temperature parameters. These replicas are allowed to swap their states after an update sweep according to a chosen criterion. One of the most used is the Metropolis-Hastings, which defines $p_{\text{SWAP}} = \min(1, e^{-\beta \Delta E})$. Another possibility is the fully deterministic swap, which does not allow unfavorable swapping. Parallel tempering leverages the swapping mechanism to employ high temperature replicas as diversification elements (sample the space effectively), and low temperature ones as intensification elements (minimize the energy effectively).

**Simulated quantum annealing.** Simulated quantum annealing (SQA) is a quantum-inspired algorithm that emulates quantum tunneling phenomena within an Ising model by incorporating a transverse field into the Hamiltonian. The Hamiltonian of the transverse Ising model is given by

$$\mathcal{H}_{\text{SQA}}(s) = - \sum_k^R \left( \sum_{i<j}^N J_{ij} s_{k,i} s_{k,j} + \sum_i^N h_i s_{k,i} \right) - J_T \sum_i^N \sum_k^R s_{k,i} s_{k+1,i}, \quad (4)$$

where the state $s$ is a $N \times R$ matrix, $R$ is the number of replicas of the system. The $i^{th}$ spin of the $k^{th}$ replica is thus identified as $s_{k,i}$. The first term of Eq. (4) is equivalent to the Ising Hamiltonian of Eq. (1) for each replica. The second term represents the transverse Hamiltonian, with $J_T$ being the intensity of the transverse field. The energy of this component is minimized if each replica is in the same state. Here, $s_{R+1,i} \equiv s_{1,i}$. The strength of the transverse field is close to zero at the beginning of the simulation, allowing the replicas to evolve relatively independently. This field smoothly increases over time until it diverges near the end of the simulation, causing the replicas to become strongly coupled and ideally collapse into one spin configuration. The formulation of the strength of the transverse coupling between replicas used in this work is given by[54]

$$J_T(n) = -J_{T0} \log\left(\tanh\left(\beta \frac{Z-n}{Z-1} G_X\right)\right), \tag{5}$$

where $J_{T0}$ is a scaling parameter, $\beta$ is the inverse temperature, $G_x$ is the maximum transverse field, and $Z$ is the total number of iterations. The input signal to the $i^{th}$ p-bit from the $k^{th}$ replica then becomes

$$I_{k,i}(n) = \beta\left(\sum_j J_{ij} s_{k,j} + h_i\right) + F_{k,i}(n), \tag{6}$$

where $J_{ij}$ is the coupling coefficient between spin $s_{k,i}$ and $s_{k,j}$, $\beta$ is the inverse temperature, $F_{k,i}(n)$ describes the coupling between neighboring replicas and is given by

$$F_{k,i}(n) = J_T(n) \times (s_{k-1,i} + s_{k+1,i}), \tag{7}$$

where $J_T(n)$ is the transverse coupling strength defined in Eq. (5).

**Maximum cut problem.** The maximum cut problem consists in finding the bipartition of a given graph that maximizes the sum of the weights of the edges between the two partitions. In other words, when considering the PIM implementation of this problem, if two p-bits that are directly interacting are in different states, the edge between the corresponding nodes of the graph can be considered cut. This means that an edge with a positive weight (if it is cut, the quality improves) is encoded as negative coupling, and vice versa.

**Acknowledgements.** This work is supported by National Research Foundation (NRF) Singapore (NRF-000214-00) and the Ministry of Education, Singapore, under Tier 2 (T2EP50123-0025). This work is also supported by the projects PRIN 2020LWPKH7 "The Italian factory of micromagnetic modeling and spintronics (IT-SPIN)", PRIN_20225YF2S4 "Magneto-Mechanical Accelerometers, Gyroscopes and Computing based on nanoscale magnetic tunnel junctions (MMAGYC)", funded by the Italian Ministry of University and Research (MUR).

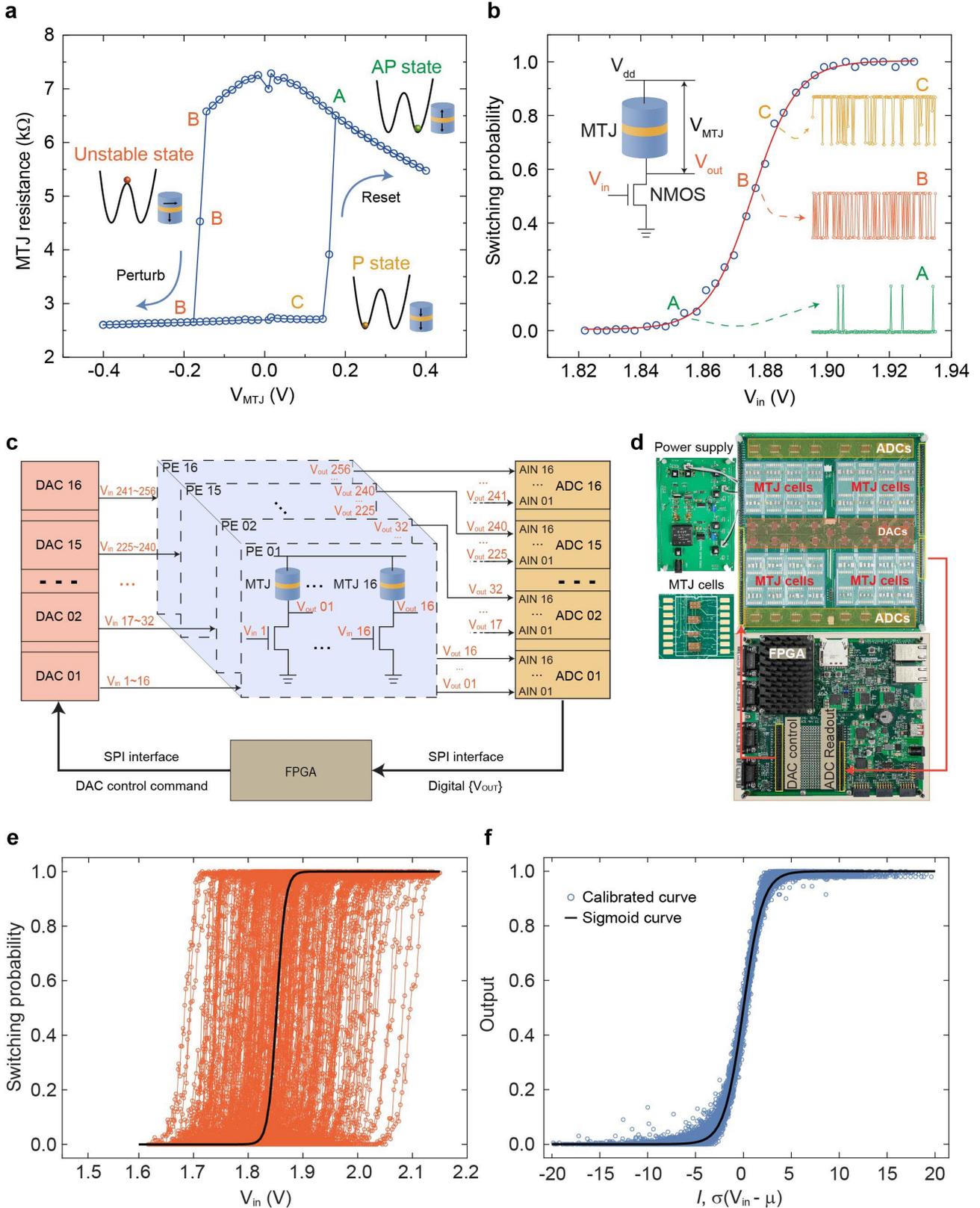

**Fig. 1| The STT-MTJ-based probabilistic Ising machine. a**, Hysteresis loop of STT-MTJ resistance versus the voltage across the MTJ ($V_{MTJ}$). The TMR ratio, ($R_{AP}$-$R_P$)/$R_P$, is 180%, where $R_{AP}$ is the antiparallel (AP) state resistance and $R_P$ is the parallel (P) state resistance. **b**, The probabilistic switching response of the STT-MTJ. The inset shows the STT-MTJ-based p-bit cell. Each data point is averaged from 10000 switching processes. Before each probabilistic switching trial, a deterministic

reset switching is performed to restore the MTJ to the AP state. The traces (A, B, and C) correspond to different input voltages ($V_{in}$) and illustrate the voltage-dependent stochastic switching behavior. Each datapoint is collected after a reset-perturb cycle. Each trace reflects controlled probabilistic switching in response to repeated identical reset-perturb pulses. **c**, System diagram of the STT-MTJ-based PIM. The system contains 16 process elements (PEs), and each PE contains 16 MTJ unit cells. There are 16 digital-to-analog converters (DACs; each 16-channel) and analog-to-digital converters (ADCs) in the system to write and read each MTJ cell. They are controlled by an FPGA through serial peripheral interface (SPI) interface. FPGA sends 24-bit command codes (DAC control command) to DACs and receives 16-bit digital representations of Vout ($V_{OUT}$) from ADCs via SPI. **d**, Photograph of the STT-MTJ-based PIM. Top left: power supply. Top right: STT-MTJ carrier printed circuit board (PCB). Bottom: FPGA board. **e**, Switching probability versus input voltage $V_{in}$ to the gate of the NMOS transistors. Black line: $P_p = 1/(1 + e^{-128.8(V_{in} - 1.852\,V)})$. **f**, The STT-MTJ switching probability curves after calibration. Black line: standard sigmoid curve.

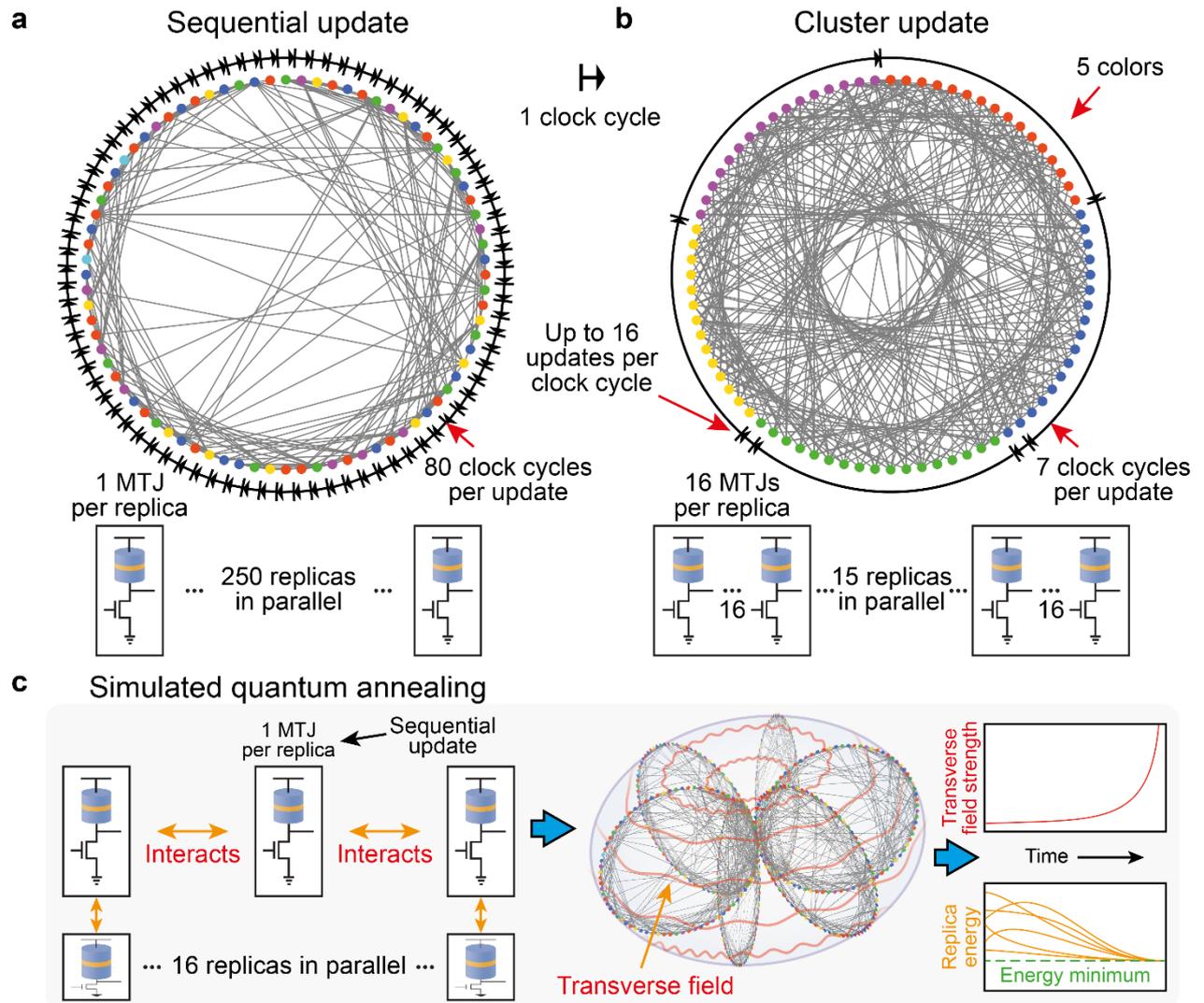

**Fig. 2| Illustration of sequential and parallel update schemes, and simulated quantum annealing.**
**a**, Sequential update scheme of the 10-bit (80 p-bits) integer factorization problem. The p-bits (represented by the colored circles) are updated one by one due to the nature of Gibbs sampling. Thus, one complete update of the graph takes 80 clock cycles. One replica contains 1 MTJ cell. With this setup, 250 replicas can be run in parallel. **b**, Cluster parallel update scheme of the 10-bit (80 p-bits) integer factorization problem. The graph is partitioned into 5 different colors using a greedy graph coloring algorithm. Each cluster is thus an independent set. Correct sampling only requires not to update adjacent p-bits at the same time, so p-bits with the same color can be updated in parallel. In our system, we divide 250 MTJs into 15 replicas with 16 MTJs per replica. 15 replicas perform the same task in parallel. If there are more than 16 p-bits with the same color, it requires more than 1 clock cycle to update the cluster. **c**, Simulated quantum annealing scheme. For SQA, sequential update is adopted. Replicas are arranged in 15 independent cyclic graphs, each with 16 replicas, and they interact with their closest neighbor through a transverse field. In an annealing schedule, the transverse field is initially negligible, but increases with time, leading to the replicas collapsing into an energy minimum.

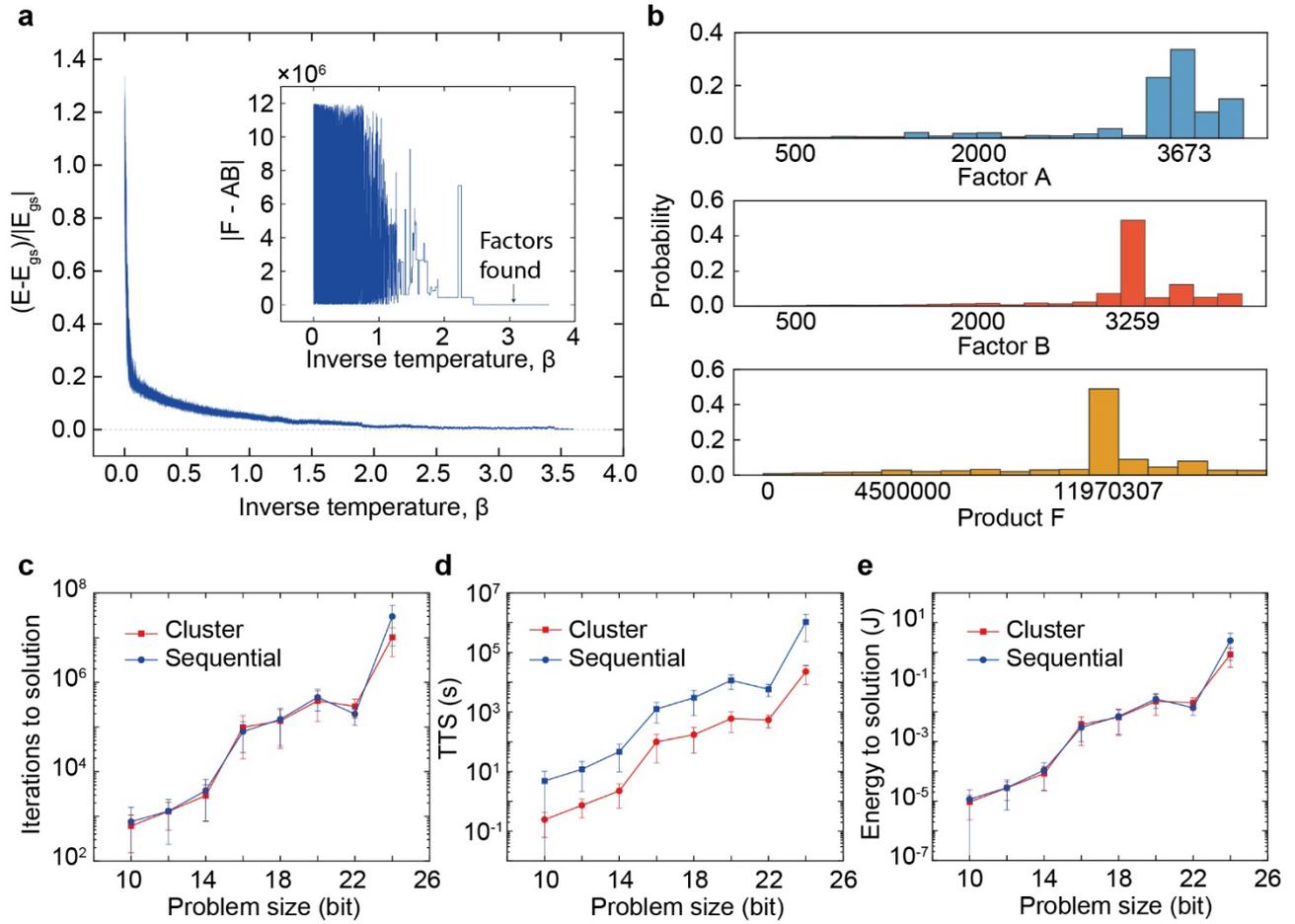

**Fig. 3| Integer factorization results. a**, Normalized energy as a function of inverse temperature. $E_{gs}$ is the ground state energy. The inset shows the solution cost $|F - AB|$ as a function of inverse temperature. **b**, Histograms showing the frequency of visiting factor *A*, factor *B* and product *F* values over the entire annealing process. **c-e**, p-bit updates per MTJ (**c**), time-to-solution (TTS) (**d**) and energy to solution (**e**) scaling with the problem size of integer factorization.

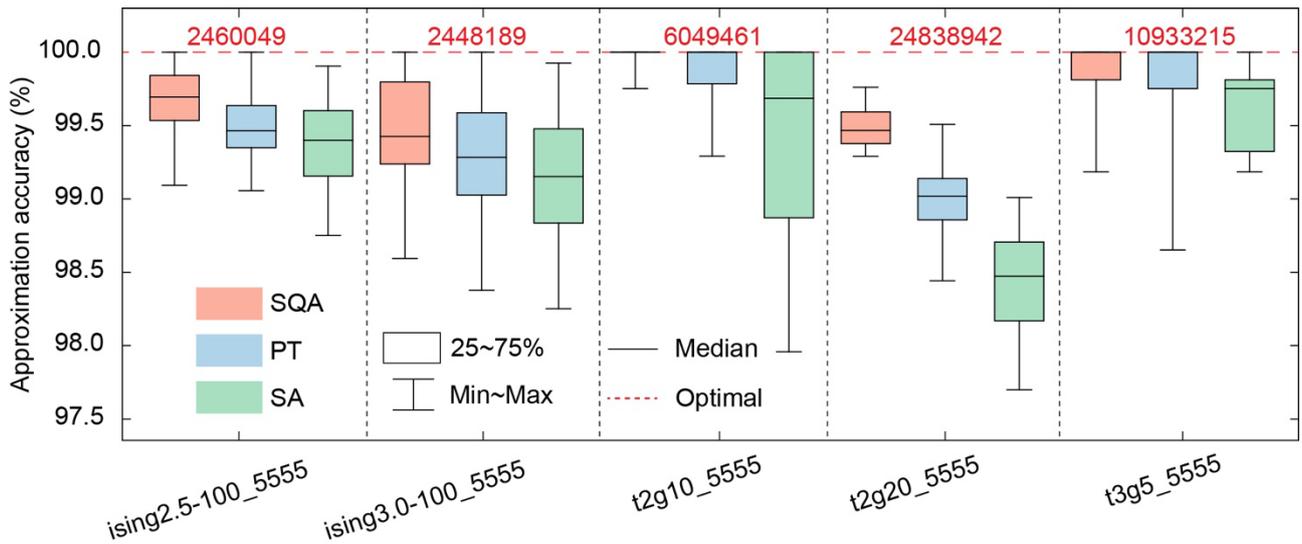

**Fig. 4| Replicated simulated annealing, parallel tempering, and simulated quantum annealing performance comparison on various Max-Cut problems.** The approximation accuracy is defined as *cut value* divided by *optimal cut value*, where a higher value represents better performance. The line splitting the box represents the median value, showing that 50% of the data lies below this value. The lower edge of the box represents the lower quartile (25%), while the upper edge represents the upper quartile (75%). The lower and upper whiskers represent the minimum and maximum approximation accuracy, respectively. The actual optimal cut value for each dataset is marked in red above the upper dotted line.

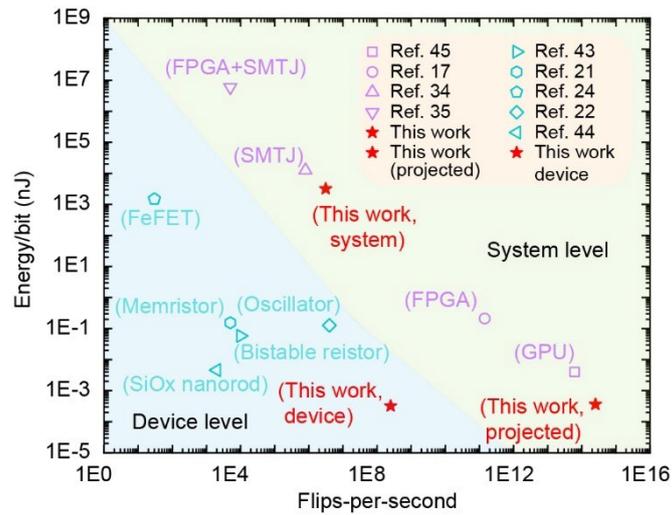

**Fig. 5 | Performance comparison of state-of-the-art PIMs.** PIMs with better performance (higher FPS and lower energy consumption) are positioned closer to the lower right corner. At the system level, the energy consumption of the entire system is evaluated, while at the device level, only the energy consumption of the main kernel is considered. GPU and FPGA based PIMs generate pseudo-random numbers, while other implementations generate true-random numbers.